\documentclass{aa}  

\usepackage{graphicx}
\usepackage{txfonts}
\usepackage{natbib} 
\bibpunct{(}{)}{;}{a}{}{,} 
\usepackage{hyperref}   
\hypersetup{colorlinks=true,linkcolor=blue,citecolor=blue,filecolor=blue,urlcolor=blue}

\usepackage{comment}
\usepackage{siunitx}

\usepackage{amsmath,amssymb}
\usepackage{mathtools}

\newcommand{\co}{^{12}{\rm C}(\alpha,\gamma)^{16}{\rm O}}

\begin{document}

\title{Impact of the uncertainties of $3 \alpha$ and $\co$ reactions on the He-burning phases of low- and intermediate-mass stars}
\subtitle{}

\author{F. Tognini \inst{1}, G. Valle \inst{1,2}, M. Dell'Omodarme \inst{1}, S. Degl'Innocenti
     \inst{1, 2}, P.G. Prada Moroni \inst{1,2}}

\titlerunning{Influence of $3 \alpha$ and $\co$ reactions rates on He-burning phases}
   \authorrunning{Tognini, F. et al.}
 \institute{
Dipartimento di Fisica ``Enrico Fermi'',
Universit\`a di Pisa, Largo Pontecorvo 3, I-56127, Pisa, Italy
\and
 INFN,
 Sezione di Pisa, Largo Pontecorvo 3, I-56127, Pisa, Italy
 }   

   \offprints{valle@df.unipi.it}

   \date{Received 11/03/2023; accepted }

  \abstract
{}
{We aim to estimate the impact on the stellar evolution of the uncertainties in the $3\alpha$ and the $\co$ reaction rates, taking into account the recent improvements in their precision.}
{ We calculated models of low- and intermediate-mass stars  for different values of $3\alpha$ and $\co$ reaction rates. The $3\alpha$ reaction rate was varied up to $\pm 24\%$ around the reference value, while  the $\co$ reaction rate was varied by up to $\pm 35\%$, taking into account different recent values for these quantities available in the literature.
The models were calculated with the FRANEC evolutionary code for two different initial chemical compositions, namely, $Y=0.246$, $Z=0.0001,$  and $Y=0.28$, $Z=0.015$
to represent different stellar populations. A $M = 0.67$~$M_{\sun}$ model was chosen as representative of the first class (halo ancient stars), while for the second composition (disk stars), the $M=1.5$~$M_{\sun}$ and $M=2.5$~$M_{\sun}$ models were considered.
The impact of $3\alpha$ and $\co$ reaction rates on the central He-burning lifetime and the asymptotic giant branch (AGB) lifetime, as well as the mass of the C/O core at the central He exhaustion and the internal C and O abundances, was investigated.
}
{A variation of the $\co$ reaction rates within its nominal error resulted in  marginal differences in the analysed features among the three considered stellar masses, except for the C/O abundances. 
The central He-burning lifetime changed by less than 4\%,
while the AGB lifetime was affected only at the 1\% level. The internal C and O abundances showed greater variation, with a change of about 15\%.
The uncertainty in the $3\alpha$ reaction rate mainly influences the C and O central abundances (up to 10\%) for all the models
considered, and the AGB lifetime for intermediate mass stars (up to 5\%). Most of the investigated features were affected by less than 2\%. }
{The current uncertainty in the explored reaction rates has a negligible effect on the predicted evolutionary time scale with respect to other uncertainty sources.
On the other hand, the variability in the chemical profile left at the end of the shell He-burning phase is still relevant. We also checked that there is no interaction between the effects of the two reaction rates,  as would be
expected in the case of small perturbations.} 


\maketitle

\section{Introduction}

Nuclear reactions are the primary source of energy production in stars and determine the evolution of a star's chemical composition. Since most nuclear reactions occur in stellar interiors at lower energies than those accessible in the laboratory, their cross-sections are generally extrapolated from the data obtained at higher energies. Thus, depending on the nuclear reaction and the characteristics of the involved nuclei, the rate can be affected by a non-negligible uncertainty, which influences the evolutionary features.

In advanced stellar evolutionary phases, starting from central He ignition throughout the central and shell He-burning phases, 
the $3\alpha$ and $\co$ reactions  compete for energy production. Therefore, they directly affect the evolutionary characteristics of stars, such as the central and shell He-burning lifetimes, the chemical profile of the star, and the CO core mass at central He exhaustion and at the end of the shell He-burning phase. Furthermore, for low- and intermediate-mass stars, a stellar chemical profile variation also alters the subsequent evolutionary characteristics, particularly with respect to the white dwarf (WD) cooling time \citep{pradamoroni2002, 2003Straniero, pradamoroni2007, 2010SalarisWD, Althaus2010}, as well as impacting WD g-mode pulsation periods \citep{Degeronimo2017, Chidester2022}.

Due to their importance for He-burning and later stages, the $3\alpha$ and $\co$ reactions have been studied throughout the years to obtain more precise estimates of their rates.
The $3\alpha$ reaction rate is dominated by the presence of an excited state with $J^\pi = 0^+$ at an energy of about $7.6$~MeV, which increases the capture process by a factor of 10-100 million \citep{Fynbo2014}. One of the most frequently used rates in evolutionary codes was calculated by the NACRE collaboration \citep{1999NACRE} with an uncertainty of about $25\%$ at He-burning temperatures. More recently, the rate estimate was refined by \citet{Fynbo2005}, with an error of about $12\%$ at central He-burning temperature ($\approx 10^8$ K). This $3\alpha$ value generally agrees with other estimates of the rate, although  \citet{2020HoyleKibedi} recently reported a value that differs from current estimates by about $34\%$.  

On the other hand, determining the $\co$ rate is far more complex. Because of the compound structure of $^{16}{\rm O}$, the astrophysical $S$ factor is dominated by broad resonances which interfere with each other and with few narrow resonances superposed upon them \citep{2017deBoer}. Thus, the estimated cross-section for this reaction has always been characterised in the past by a great uncertainty, which, however, decreased during the years. The cross-section error decreased from a factor of 2 at the end of the 1990s \citep{1985CF, 1988CF} down to the $35\%$ figure reported by \citet{2002Kunz}, and then $30\%$ by \citet{Hammer2005}. A recent estimate of the $\co$ was  proposed by \citet{2017deBoer}, with an associated uncertainty of about $20\%$ at central He-burning temperature. 

A series of studies has provided a discussion of the impact  of the uncertainties in the $3\alpha$ and the $\co$ reaction rates on the stellar
evolution over the years \citep[e.g.][]{1985CF, 2001Cassisi, 2003Straniero, 2005Weiss, Dotter2009, 2013Valle, 2016Fields,  Degeronimo2017, Pepper2022}.
The aim of the present paper is to investigate how the recent improvements in the precision of the reaction rates  affect the results reported in the existing literature. The effects on the evolutionary characteristics of the $3\alpha$ and the $\co$ reaction rate changes, within the current uncertainties, were analysed separately for each reaction. We also discuss a set of results obtained by varying simultaneously the two rates at their upper and lower boundaries to investigate the presence of interactions between the effects of the two reaction rates.

\section{Methods}\label{sec:methods}

To analyse the effects of the present uncertainties on the studied reactions, models of low and intermediate mass stars were calculated for different values of $3\alpha$ and $\co$ reaction rates.

As a reference, we adopted the $3\alpha$ reaction rate by \citet{Fynbo2005} and the $\co$ reaction rate by \citet{2017deBoer}. Additional models were computed with the analysed reaction rates varied within their errors. The $3\alpha$ reaction rate was varied up to $\pm 24\%, $  which is about the double of the uncertainty by \citet{Fynbo2005}; the \citet{2020HoyleKibedi} $+34\%$ variation has also been taken into account. The $\co$ reaction rate was varied by up to $\pm 35\%$, the uncertainty associated with the \citet{2002Kunz} estimate, which is widely used in the literature. This choice is dictated by the fact that the recent high precision data on the $\co$ cross-section by \citet{2017deBoer} are still somehow debated in the literature \citep[see][for more details]{Smith2021C12}. The values of the reaction rates  were changed one by one since, for a small variation in the input physics, we do not expect correlation effects on the model results, as previously discussed in the literature \citep[e.g.][]{2013Valle}. However, we explicitly tested this assumption by performing computations that simultaneously varied  the two rates at their upper and lower boundaries. The results, presented in Appendix~\ref{app:linear}, support this claim. Therefore, any combination of the joint impact of the two reaction rate uncertainties can be obtained by a simple sum of the effects of the individual rates.

The models were calculated with the FRANEC evolutionary code in the same configuration as in the paper by \citet{DellomodarmeFRANEC}, which includes a complete description of the adopted input physics. Here, we discuss only the input relevant to a comparison with recent results.  

Models were computed without core overshooting, adopting the Schwarzschild criterion to define convective boundaries; furthermore, instantaneous mixing was assumed in convective regions.  Semiconvection during the
central He-burning phase was treated following the numerical scheme
described in \citet{castellani1985}. Breathing pulses were suppressed
\citep{2001Cassisi, Constantino2017} following the procedure
adopted in \citet{DellomodarmeFRANEC}. 

H-burning reaction rates were obtained from the NACRE database \citep{1999NACRE} with the exception of the $pp$ and the  $^{14}{\rm N}(p,\gamma)^{15}{\rm O}$ reaction rates, which are from \citet{Marcuccipp2019} and the LUNA collaboration \citep{Imbriani2005}, respectively.

Models were calculated for two different initial chemical compositions. The first one ($Y=0.246$ and $Z=0.0001$) is typical of ancient low mass stars populating the halo of our Galaxy, while the other ($Y=0.28$ and $Z=0.015$) characterises stars in the Galactic disk. Different stellar ages and mass ranges were considered in the two scenarios.
A star with $M = 0.67$~$M_{\sun}$ during the He-burning phases, with a $0.8$~ $M_{\sun}$ mass star as a progenitor, was chosen to represent ancient stars with a halo chemical composition. Regarding intermediate mass stars, $M=1.5$~$M_{\sun}$ and $M=2.5$~$M_{\sun}$ stars with  disk chemical composition have been taken as reference. The $2.5$~$M_{\sun}$ star is assumed to be in the typical mass range of progenitors of $0.55-0.6$~$M_{\sun}$ white dwarfs, which are at the peak of the field WD mass distribution \citep{pradamoroni2007, evolution_of_stars}.

\section{Effect of the $\co$ reaction rate uncertainties}
\label{sec: effetto_c12}

The $\co$ reaction is fundamental for stellar nucleosynthesis and evolution during the He-burning phases; unlike the $3\alpha$-burning, the $\co$ reaction only affects the evolutionary characteristics after the central He ignition.

The analysis focuses on the following: the central He-burning lifetime ($t_{\rm HB}$), defined as the time from the onset of He-burning until when the central He abundance drops below $10^{-6}$; the asymptotic giant branch (AGB) lifetime ($t_{\rm AGB}$), defined as the time from the central He exhaustion and the first thermal pulse; the central C and O abundances at the central He exhaustion. Central carbon and oxygen abundances, which do not change during the AGB evolution, 
are particularly relevant because they also affect the white dwarf evolution,  including
 their pulsation properties \citep{Degeronimo2017, Chidester2022}. To explore the impact on the stellar evolution of such a variation, we also investigated the chemical profile at the first thermal pulse.

The results are shown in Table~\ref{tab:co} for the three different masses discussed in Sect.~\ref{sec:methods}. Overall, only marginal differences were detected in the relative impact of the reaction rate variations among the three considered stellar masses. 

Regarding the He-burning lifetime, the increase in the $\co$ reaction rate enhances $t_{\rm HB}$ by few percents. This is expected because $3\alpha$ and $\co$ are the only reactions which contribute to the energy production in the core during the central He-burning phase. In fact, the total stellar energy production is fixed to balance the stellar radiative losses from the surface, so decreasing or increasing one of the two rates influences the relative efficiency of the two reactions. The $3\alpha$ burns three He nuclei per reaction, while for each $\co$ reaction, only one $\alpha$ particle is consumed. At fixed stellar luminosity, an increase in the $\co$ cross-section enhances the relative contribution of the reaction to the energy production. This reduces the He consumption rate, while more carbon is burned as a result; thus, the central He-burning lifetime increases. 
Overall, the variation induced by the $\co$ uncertainty in $t_{\rm HB}$ is quite negligible; namely, it s lower than other uncertainty sources, as we discuss in Sect.~\ref{sec:conclusions}.

\begin{table*}
\begin{center}
\caption{Impact of the uncertainty in the $\co$ reaction rate for the three considered stellar masses.}
\label{tab:co}
\begin{tabular}{lccccccc} 
\hline\hline
\multicolumn{8}{c}{$M= 0.67$ $M_{\sun}$, $Y=0.246$, $Z=0.0001$} \\
  & Rate $+35\%$ & Rate $+28\%$ & Rate $+20\%$ & Standard rate & Rate $-20\%$ & Rate $-28\%$ & Rate $-35\%$\\ 
  \hline
$t_{\rm HB}$ [Myr] & 101.2 & 100.8 & 100.1  & 98.4  & 96.4 & 95.4 & 94.8 \\ 
  & $+2.8\%$  & $+2.4\%$ & $+1.8\%$ &  & $-2.0\%$ & $-3.0\%$ & $-3.6\%$ \\
 \hline
 $X_{^{12}{\rm C}}$ & 0.292 & 0.309 & 0.329 & 0.384 & 0.453 & 0.482 & 0.514
 \\ 
   & $-22\%$ & $-18\%$ & $-13\%$ &  & $+18\%$ & $+26\%$ & $+34\%$ \\
 \hline
$X_{^{16}{\rm O}}$ & 0.708 & 0.691 & 0.671 & 0.616 & 0.547 & 0.517 & 0.485 \\ 
 & $+15\%$ & $+12\%$ & $+9\%$ & & $-11\%$ & $-16\%$ & $-21\%$ \\
 \hline
 $t_{\rm AGB}$ [Myr] & 12.35  & 12.37 & 12.38 & 12.52 & 12.57 & 12.62 & 12.62 \\ 
  & $-1.4\%$ & $-1.2\%$ & $-1.1\%$ & & $+0.4\%$ & $+0.8\%$ & $+0.8\%$ \\
 \hline 
 \multicolumn{8}{c}{$M= 1.5$ $M_{\sun}$, $Y=0.28$, $Z=0.015$} \\
 & Rate $+35\%$ & Rate $+28\%$ & Rate $+20\%$  & Standard rate & Rate $-20\%$ & Rate $-28\%$ & Rate $-35\%$\\ 
 \hline
$t_{\rm HB}$ [Myr] & 120.3 & 119.9 & 119.3 & 117.4 & 115.2 & 114.2 & 113.2 \\ 
& $+2.9\%$ & $+2.1\%$ & $+1.6\%$ & & $-1.8\%$ & $-3.2\%$ & $-3.6\%$ \\
\hline
$X_{^{12}{\rm C}}$  & 0.289 & 0.304 & 0.323 & 0.377 & 0.444 & 0.475  & 0.504
 \\
& $-23\%$ & $-19\%$ & $-14\%$ & & $+18\%$ & $+26\%$ & $+34\%$ \\
\hline
$X_{^{16}{\rm O}}$ & 0.691  & 0.676 & 0.657 & 0.603 & 0.536 & 0.505 & 0.475 \\ 
& $+15\%$ & $+12\%$ & $+10\%$ & & $-11\%$ & $-16\%$ & $-24\%$ \\
\hline
$t_{\rm AGB}$ [Myr] & 12.99 & 13.03 & 13.06 & 13.17 & 13.38 & 13.46 & 13.55 \\ 
 & $-1.4\%$ & $-1.1\%$ & $-0.8\%$ & & $+1.5\%$ & $+2.2\%$ & $+2.8\%$ \\
\hline
 \multicolumn{8}{c}{$M= 2.5$ $M_{\sun}$, $Y=0.28$, $Z=0.015$} \\
  & Rate $+35\%$ & Rate $+28\%$ & Rate $+20\%$ & Standard rate  & Rate $-20\%$ & Rate $-28\%$ & Rate $-35\%$\\ 
 \hline
$t_{\rm HB}$ [Myr] & 244.5 & 243.8 & 242.9 & 240.2 & 237.1 & 235.9 & 234.3\\ 
& $+1.8\%$ & $+1.5\%$ & $+1.2\%$ & & $-1.3\%$ & $-1.8\%$ & $-2.4\%$\\
 \hline
 $X_{^{12}{\rm C}}$ & 0.259 & 0.274 & 0.293 & 0.348 & 0.415 & 0.444  & 0.475
 \\
& $-25.6\%$  & $-21\%$ & $-16\%$ & & $+20\%$ & $+27\%$ & $+36\%$\\
 \hline
$X_{^{16}{\rm O}}$ & 0.721 & 0.706  & 0.687 & 0.631 & 0.565 & 0.536 & 0.505  \\ 
& $+14\%$ & $+12\%$ & $+9\%$ & & $-10\%$ & $-15\%$ &  $-20\%$\\
 \hline
 $t_{\rm AGB}$ [Myr] & 18.25 & 18.27 & 18.33 & 18.95  & 19.34 & 19.37 & 19.52 \\ 
&  $-3.6\%$ & $-3.5\%$ & $-3.0\%$ & & $+2.0\%$ & $+2.2\%$ &  $+3.0\%$\\
 \hline
\end{tabular}
\tablefoot{
 The considered evolutionary characteristics are: central He-burning lifetime ($t_{\rm HB}$), carbon and oxygen central abundances ($X_{^{12}{\rm C}}$, $X_{^{16}{\rm O}}$) at He exhaustion and evolutionary time during the AGB phase ($t_{\rm AGB}$). The percentage variation from the value obtained with the standard rate is also reported.}
\end{center}
\end{table*}

\begin{figure*}
   \centering
    \includegraphics[width=160mm]{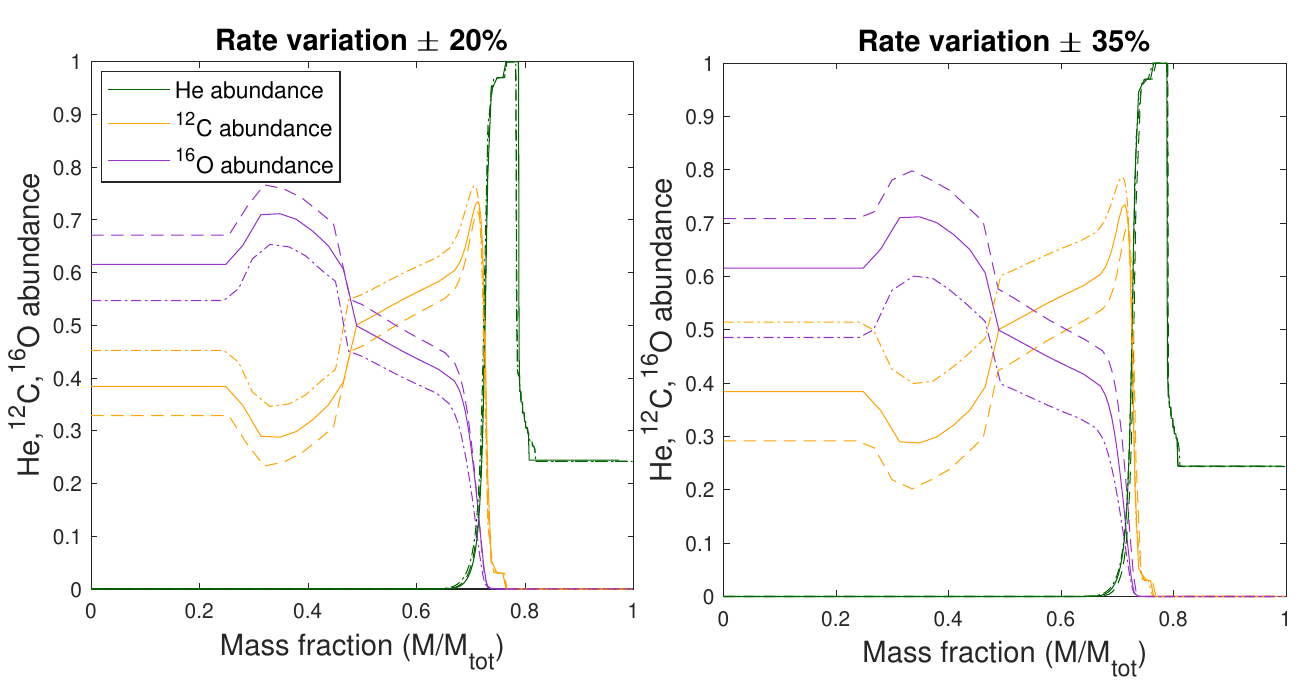}
    \caption{ Profile of $^4$He, $^{12}$C and $^{16}$O abundances at the first thermal pulse as a function of mass fraction. {\it Left}: Profile obtained by increasing (dashed lines) and decreasing (dot dashed lines) the $\co$ reaction rate by $20\%$. 
    The continuous lines correspond to the data obtained with the standard rate. The green lines represent the abundance of $^4$He, the orange lines the one of $^{12}$C, and the purple lines the one of $^{16}$O.
     {\it Right}: Same as in the left panel but for a rate variation of $\pm35\%$.}
    \label{fig:abbondanzeC12}
\end{figure*}

A change in the $\co$ reaction rate also impacts the chemical composition of the whole region inside the He shell and of the region affected by semi-convective mixing at central He exhaustion and at the first thermal pulse.
At the central He exhaustion, the abundances of carbon and oxygen in the core and in the semi-convective region are significantly modified, while the remaining external parts of the previous He core remain unchanged. As expected, increasing (decreasing) the $\co$ reaction rate decreases (increases) the remaining $^{12}{\rm C}$ abundance. The greater the variation in the rate, the higher the change in the abundances. The carbon and oxygen central abundances are significantly affected by the changes in the $\co$ reaction rate, with a maximum variation between $-22\%$ and $+34\%$ in ${^{12}{\rm C}}$ and between $+15\%$ and $-21\%$ in ${^{16}{\rm O}}$.
Then the He-burning continues in shell during the AGB phase. The effect of different $\co$ rates on the chemical profile at the first thermal pulse can be seen in Fig.~\ref{fig:abbondanzeC12}.
This modification has a relevant impact on stellar nucleosynthesis and also mildly affects the AGB evolution. Moreover, the variation of the chemical composition influences the cooling time of the stars in the following white dwarf phase \citep[see e.g.][]{pradamoroni2002, 2003Straniero, pradamoroni2007,  2010SalarisWD, Salaris2013, Salaris2022}; a precise knowledge of the chemical
profile of the progenitor star is of fundamental importance to the understanding of the thermal capacity of the stellar matter and, thus, how cooling carries on. This has, in turn, consequences on the globular cluster or galactic disk and halo age determinations through the comparison between theory and observation for the white dwarf cooling sequence \citep{pradamoroni2002, 2003Straniero, pradamoroni2007,  2010SalarisWD, Hansen2013, GarciaBerro2014, Kilic2019, Prisegen2021}.

Finally, $t_{\rm AGB}$ is only mildly affected, at the level of few percents. As discussed in \citet{BossiniConvezione} and \citet{2014Cassisi}, an increase in the extension of the central He-burning convective region, and thus of the $^{12}{\rm C}/^{16}{\rm O}$ core mass at the central He exhaustion, reduces the AGB lifetime, since the He shell forms closer to the H shell. Increasing the $\co$ reaction rate enhances, even if very slightly, the $^{12}{\rm C}/^{16}{\rm O}$ core mass  at the central He exhaustion and thus the AGB lifetime is slightly reduced.
Other quantities such as the luminosity of the AGB clump or the $^{12}{\rm C}/^{16}{\rm O}$ core mass at the first thermal pulse are practically unaffected by a change within the present errors of the $\co$ reaction rate, with variations lower than $1\%$.

The effects of varying the $\co$ reaction rates have been extensively studied in the literature, mainly when the related uncertainty was quite large. As an example, the error of the $\co$ reaction rate was of a factor of 2 at the beginning of the 90s of the last century \citep{1985CF, 1988CF}. From the results reported in \citet{2001Cassisi} and \citet{2001Imbriani}, for stars of $0.75-0.8$ $M_{\sun}$, one can see that a change of this order of magnitude in the reaction rate produces a variation in $t_{\rm HB}$ by about $10\%$, while the change in  $^{12}$C and $^{16}$O abundances is about $60\%$ and $40\%$, respectively, in agreement with our results. 
The data obtained for the $2.5$ $M_{\sun}$ are also consistent with the present literature. From the results obtained by \citet{2016Fields} and \citet{2003Straniero}, it is apparent that our data follow the same trend with the $\co$ reaction rate, with comparable percentage variations.
A relevant comparison is with the results recently reported by \citet{Pepper2022} who investigate the impact of different $\co$ reaction rates in several stellar evolutionary phases. Their analysis adopts MESA tracks, computed under different assumptions about diffusion, treatment of semi-convection, and suppression of breathing pulses with respect to those adopted in our tracks. Among the explored rates, \citet{Pepper2022} adopted NACRE values, assuming an uncertainty of 32\%, very similar to the 35\% adopted here. Their mass ranges overlap with our $M = 1.5$ $M_{\sun}$ and $M = 2.5$ $M_{\sun}$ models, allowing for a direct comparison. It is however important to consider that the models were computed assuming a different rate functional form, because we adopted \citet{2017deBoer} results as our reference; thus, at different metallicity, the impact of these differences cannot be correctly accounted for.

Regarding the C/O ratio at the central He exhaustion \citet{Pepper2022} report 0.520, varying in the range (0.310; 0.977), and 0.492 (0.289; 0.935) for $M$ = 1.5 and 2.5 $M_{\sun}$, respectively. For comparison, we obtained 0.625 (0.418; 1.06) and 0.551 (0.359; 0.941). While the results are quite similar for $M = 2.5$ $M_{\sun}$, there is a non-negligible offset for the 1.5 $M_{\sun}$ model. However, the spanned range is similar.

\citet{Pepper2022} also investigated on the difference in $t_{\rm HB}$ reporting  variations within the ranges of (-6.6; 3.8) Myr and (-6.5; 3.5) Myr for 1.5 and 2.5 $M_{\sun}$, respectively. Our results for the two considered stellar masses are in the range (-4.2; 2.9) Myr and (-5.9; 4.3) Myr. Also, in this case, the results show a better agreement for the more massive star, the difference in the range width for the lightest one being about 50\%.

A change in the $\co$ reaction rate also affects the theoretical estimate of two observational parameters, the $R$ and $R2$ parameters. The former is defined as
\begin{equation}
\label{eq: parametro_R}
    R=\frac{N_{\rm HB}}{N_{\rm RGB}} = \frac{t_{\rm HB}}{t_{\rm RGB}},
\end{equation}
where $N_{\rm HB}$ is the number of stars in the horizontal branch and $N_{\rm RGB}$ is the number of stars in the red giant branch (RGB) with a luminosity greater than the zero-age horizontal branch (ZAHB). As described in \citet{Cassisi2003}, it is possible to estimate the $R$ parameter via theoretical models, as in Eq.~(\ref{eq: parametro_R}), where $t_{\rm HB}$ is the He-burning lifetime from the ZAHB to central He exhaustion, while $t_{\rm RGB}$ is the life time of the star during the RGB from the point in which the luminosity is higher than the ZAHB; for the ZAHB luminosity, we can assume the luminosity of its horizontal portion. The $R$ parameter is the most important indirect indicator for the original He abundance in globular cluster stars; however to appropriately use this parameter, a precise evaluation of the associated uncertainty is mandatory. 

The $R2$ parameter is defined as:
\begin{equation}
\label{eq: R2}
    R2=\frac{N_{\rm AGB}}{N_{\rm HB}} = \frac{t_{\rm AGB}}{t_{\rm HB}},
\end{equation}
that is as the ratio between the number of stars in the AGB phase ($N_{\rm AGB}$) and in the horizontal branch phase ($N_{\rm HB}$). As described in \citet{BuonannoR2}, the $R2$ parameter can be calculated in theoretical models with the ratio between the AGB lifetime and the HB lifetime. Thus, its theoretical estimate is affected by the changes in both $t_{\rm HB}$ and $t_{\rm AGB}$. 

A precise knowledge of the uncertainty related to this parameter is fundamental because it is particularly useful to obtain information on mixing processes during the central He-burning phase. For example, the study of the $R2$ parameter was used to determine if breathing pulses are numerical artifacts in evolutionary codes or a real process during stellar evolution \citep[see also][]{BuonannoR2, CassisiR}.

Table~\ref{tab: R_C12} reports the theoretical values of the $R$ and $R2$ parameters obtained for a $0.67$~$M_{\sun}$ for the selected $\co$ reaction rate values. The effect is only few percents, which is significantly lower than the current observational uncertainty that are affected by the errors in star counts (about 30\%; for more details see \citealt{CostantinoConvezione1} or \citealt{CassisiR}). Thus, at the present level of precision, the $\co$ reaction rate no longer constitutes a relevant error source for the theoretical estimation of $R$ and $R2$ parameters.

\begin{table*}
\begin{center}
\caption{$R$ and $R2$ parameters for different $\co$ reaction rates. The percentage variation with respect to the standard rate is also shown.}
\label{tab: R_C12}
\begin{tabular}{ lccccccc} 
\hline\hline
 $0.67$ $M_{\sun}$ & Rate $+35\%$ & Rate $+28\%$ & Rate $+20\%$ & Standard rate & Rate $-20\%$ & Rate $-28\%$ & Rate $-35\%$\\ 
 \hline
 $R$ parameter & 0.976  & 0.972 & 0.965 & 0.949 & 0.929 & 0.920 & 0.914 \\ 
  & $+2.8\%$ & $+2.4\%$ & $+1.8\%$ & & $-2.0\%$ & $-3.0\%$ & $-3.6\%$ \\
 \hline
$R2$ parameter & 0.1229  & 0.1231 & 0.1237 & 0.1273 & 0.1305 & 0.1323 & 0.1326 \\ 
& $-3.5\%$ & $-3.3\%$ & $-2.8\%$ & & $+2.5\%$ & $+4.0\%$ & $+4.5\%$ \\
 \hline 
\end{tabular}
\end{center}

\end{table*}

\section{Effect of the $3\alpha$ reaction rate uncertainties}
\label{sec:effetto3a}

\begin{figure}
   \centering
    \includegraphics[width=90mm]{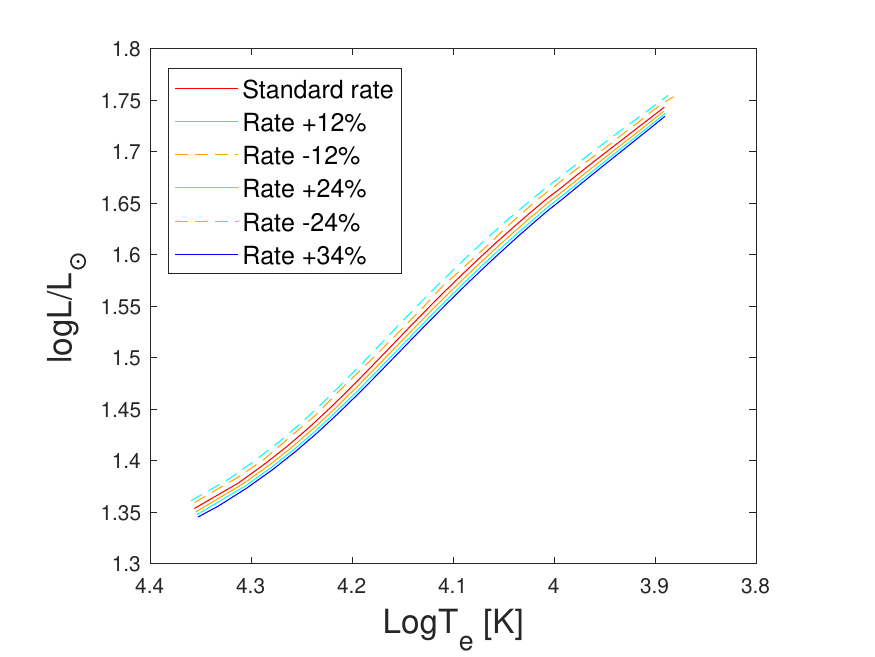}
    \caption{Zero-age horizontal branch for different values of the $3\alpha$ reaction rate. }
    \label{fig: ZAHB_3a}
\end{figure}

Stellar models were computed assuming $3\alpha$ reaction rate variations of $\pm 12\%$ (the present uncertainty for the $3\alpha$ reaction rate from \citealt{Fynbo2005, Fynbo2014}), $\pm 24\%$ ($2\sigma$ variation) and $+34\%$. The latter is related to the data discussed in \citet{2020HoyleKibedi}.

Thus, in addition to the evolutionary characteristics investigated for the $\co$ reaction rate, we studied also the effect of the $3\alpha$ on the He core mass, stellar age, and luminosity at the RGB tip, and the ZAHB characteristics. The results are reported in Table~\ref{tab:3a} for the three selected models.

\begin{table*}
\begin{center}
\caption{
As in Table~\ref{tab:co}, but for different values of the $3\alpha$ reaction rate. }
\label{tab:3a}
\begin{tabular}{ lcccccc} 
\hline\hline
\multicolumn{7}{c}{$M= 0.67$ $M_{\sun}$, $Y=0.246$, $Z=0.0001$} \\
& Rate $+34\%$ & Rate $+24\%$ & Rate $+12\%$ & Standard rate & Rate $-12\%$ & Rate $-24\%$  \\ 
\hline
$M_{\rm He}$ [$M_{\sun}$] &  0.5084 & 0.5089 &  0.5098 &   0.5108 &  0.5118 & 0.5131\\ 
& $-0.5\%$ & $-0.4\%$ & $-0.2\%$ & & $+0.2\%$  & $+0.4\%$ \\
 \hline
 $\log{L_{\rm ZAHB}/L_{\sun}}$ & 1.6880 & 1.6905 & 1.6943 & 1.6986 & 1.7032 &  1.7085 \\ 
  & $-0.6\%$ & $-0.5\%$ & $-0.2\%$ &  & $+0.2\%$ & $+0.6\%$ \\
 \hline
 $t_{\rm HB}$ [Myr] & 97.9 & 98.0 & 98.2 & 98.4 & 99.8 & 98.8\\ 
 & $-0.5\%$ & $-0.4\%$ & $-0.2\%$ & & $+1.4\%$ & $+0.4\%$ \\
 \hline
 $X_{^{12}{\rm C}}$  & 0.440 & 0.426 & 0.405 & 0.384 & 0.360 & 0.331\\ 
& $+15\%$ & $+11\%$ & $+5\%$ & & $-6\%$ & $-14\%$ \\
 \hline
 $X_{^{16}{\rm O}}$ & 0.560 & 0.574 & 0.595 & 0.616 & 0.640 & 0.669\\ 
& $-9\%$ & $-7\%$ & $-3\%$ & & $+4\%$ & $+8\%$ \\
 \hline
 $t_{\rm AGB}$ [Myr] & 12.66 & 12.63  & 12.60 & 12.52 & 12.50 & 12.38 \\ 
 & $+1.1\%$ & $+0.9\%$ & $+0.6\%$ & & $-0.2\%$ & $-1.1\%$\\
 \hline
  \multicolumn{7}{c}{$M= 1.5$ $M_{\sun}$, $Y=0.28$, $Z=0.015$} \\
 & Rate $+34\%$ & Rate $+24\%$ & Rate $+12\%$ & Original rate & Rate $-12\%$ & Rate $-24\%$ \\ 
 \hline
$t_{\rm HB}$ [Myr] & 117.1 & 117.2 & 117.3 & 117.4 & 117.5 & 117.6\\ 
 & $-0.3\%$ & $-0.2\%$ & $-0.1\%$ & & $+0.1\%$  & $+0.2\%$ \\
 \hline
 $X_{^{12}{\rm C}}$ & 0.432 & 0.418 & 0.399 & 0.377 & 0.354 & 0.327\\ 
& $+14\%$ & $+11\%$ & $+6\%$ & & $-6\%$  & $-13\%$ \\
\hline
$X_{^{16}{\rm O}}$ & 0.548 & 0.562 & 0.581 & 0.603 & 0.626 & 0.653\\ 
& $-9\%$ & $-7\%$ & $-4\%$ & & $+4\%$  & $+8\%$ \\
 \hline
 $t_{\rm AGB}$ [Myr] & 13.63 & 13.52 & 13.29 & 13.17 & 12.97 & 12.80 \\ 
 & $+3.8\%$ & $+3\%$ & $+1.3\%$ & & $-1.1\%$  & $-2.5\%$ \\
 \hline
\multicolumn{7}{c}{$M= 2.5$ $M_{\sun}$, $Y=0.28$, $Z=0.015$} \\
 & Rate $+34\%$ & Rate $+24\%$ & Rate $+12\%$ & Standard rate & Rate $-12\%$ & Rate $-24\%$ \\ 
 \hline
$t_{\rm HB}$ [Myr] & 237.5 & 237.8  & 238.6 & 240.2 & 240.8 & 241.4\\ 
& $-1.1\%$ & $-1.0\%$ & $-0.6\%$ & & $+0.3\%$  & $+0.5$ \\
\hline
$X_{^{12}{\rm C}}$ & 0.403 & 0.389 & 0.369 & 0.348 & 0.325 & 0.299\\ 
 & $+16\%$ & $+12\%$ & $+6\%$ & & $-7\%$  & $-14\%$ \\
\hline
$X_{^{16}{\rm O}}$  & 0.577 & 0.591 & 0.611 & 0.631 & 0.655 & 0.682\\
 & $-8.6\%$ & $-6.4\%$ & $-3.3\%$ & & $+4.0\%$  & $+8.0\%$ \\
 \hline
 $t_{\rm AGB}$ [Myr] & 19.99 & 19.78 & 19.47 & 18.95 & 18.69 & 18.42\\ 
& $+5.7\%$ & $+4.6\%$ & $+3.0\%$ & & $-1.1\%$  & $-2.5\%$ \\
 \hline
\end{tabular}
\end{center}
\end{table*}

At the end of the RGB evolutionary phase low mass stars ignite He in an electron degenerate core, at the RGB tip. The luminosity at He ignition is directly related to the He core mass, which depends on the temperature needed to activate the $3\alpha$ reaction; thus, it depends on the $3\alpha$ reaction rate itself. An increase (decrease) of $3\alpha$ reaction rate corresponds to a slightly lower (greater) temperature required to activate the reaction and to a reduced (enhanced) He core mass at the RGB tip \citep[see e.g.][]{2013Valle}.

As can be seen in Table~\ref{tab:3a} the change in $M_{\rm He}$ is quite negligible, with a maximum variation of between $-0.5\%$ and $0.4\%$. This produces a negligible change in the luminosity of the star at the RGB tip and at the ZAHB (see Fig.~\ref{fig: ZAHB_3a}), with a maximum change below $0.3\%$. Furthermore, the stellar age at the RGB tip is unchanged, with a variation of about $0.001\%$, in agreement with the results by \citet{2013Valle}.

The effect on the central He-burning lifetime of the $3\alpha$ reaction variation is quite small, with a maximum of about $1\%$ for all the selected values, lower than the effect due to changes of the $\co$ rate. For the $0.67$~$M_{\sun}$ a variation of $-12\%$ in the rate reduces $t_{\rm HB}$ more than a change of $-24\%$; this is connected to the breathing pulses treatment. 
The start of breathing pulses is directly related to the He central abundance, therefore, changing the rate at which He is burned affects the time at which breathing pulses start, and their length. This change of about $1\%$ in $t_{\rm HB}$ is the typical variation associated with the $3\alpha$ reaction rate in the literature \citep[see e.g.][]{2005Weiss}.

\begin{figure*}
   \centering
    \includegraphics[width=160mm]{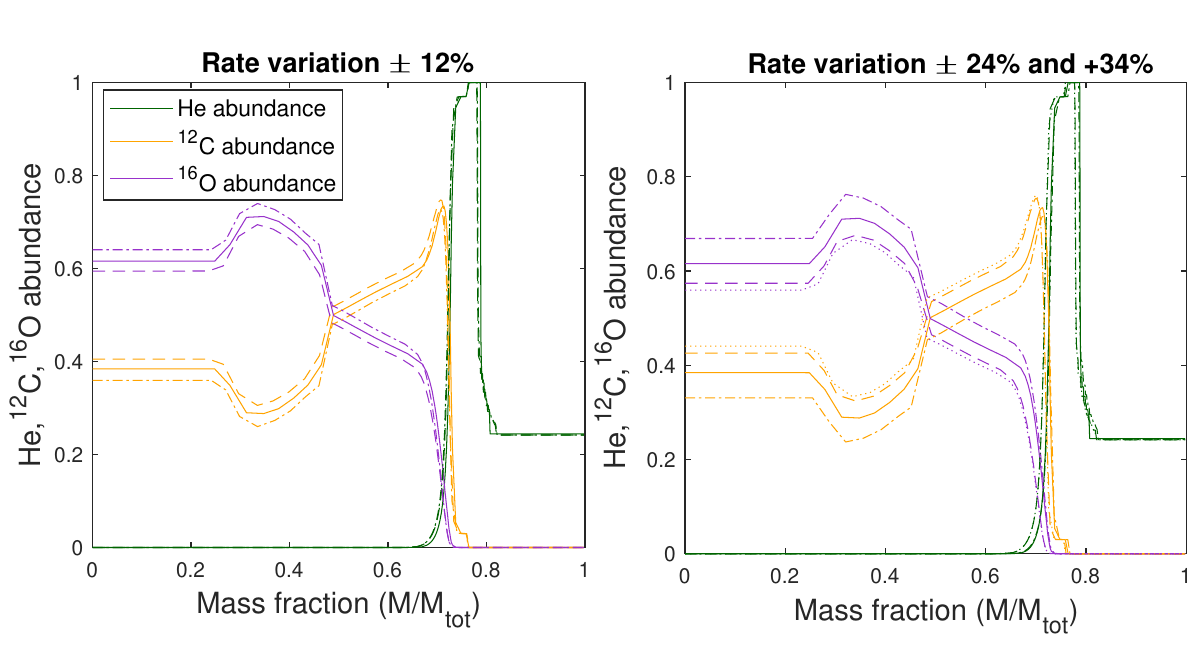}
    \caption{ Profiles of $^4$He, $^{12}$C, and $^{16}$O abundances at the first thermal pulse as a function of mass fraction. {\it Left}: Profiles obtained increasing (dashed lines) and decreasing (dot-dashed lines) the $3\alpha$ reaction rate by $12\%$. The continuous lines correspond to the data obtained with the standard rate. The green lines represent the abundance of $^4$He, the orange ones the abundance of $^{12}$C, and the purple lines the one of $^{16}$O. {it Right}: Same as in the left panel, but for a  $3\alpha$ reaction rate variation by  $\pm24\%$ and $+34\%$ (dotted lines).}
    \label{fig: abbondanze3a_1tp}
\end{figure*}

The chemical profiles of the stars at the first thermal pulse are significantly affected by variations of $3\alpha$ reaction rate, as shown in Fig.~\ref{fig: abbondanze3a_1tp}. As expected, a variation of the $3\alpha$ reaction rate influences the carbon and oxygen abundances inside the $^{12}{\rm C}/^{16}{\rm O}$ core, in the region which experienced semi-convection and all the zones affected by the He-burning shell. Increasing (decreasing) the $3\alpha$ reaction rate enhances (reduces) the $^{12}{\rm C}$ abundance while the $^{16}{\rm O}$ one lowers (grows).
The change in $^{12}{\rm C}$ and $^{16}{\rm O}$ central abundances is generally smaller (about 60\%) than the one that is due to the $\co$ rate variation. 

The $3\alpha$ reaction rate also affects the evolutionary time during the AGB phase. Increasing the $3\alpha$ reaction rate enhances $t_{\rm AGB}$ because the $^{12}{\rm C}/^{16}{\rm O}$ core lowers in mass. The change in $t_{\rm AGB}$ with the $3\alpha$ reaction rate increases with the stellar mass, with a maximum variation of about $5\%$ for the $2.5$~$M_{\sun}$ mass star. The $3\alpha$ reaction rate also affects the evolutionary time at the first thermal pulse ($t_{\rm HB} + t_{\rm AGB}$) with a change similar to the central He-burning lifetime because $t_{\rm AGB}$ is almost an order of magnitude lower than $t_{\rm HB}$.
Lastly, the $^{12}{\rm C}/^{16}{\rm O}$ core mass and the luminosity of the AGB clump are marginally affected by a change in the $3\alpha$ reaction rate, with a maximum variation lower than $1\%$.

The impact of a $3\alpha$ reaction rate variation was analysed in the literature mainly when the uncertainties were higher. The results of \citet{2005Weiss} or \citet{2013Valle} show that for low mass stars, a variation of about $20\%$ produces an almost negligible effect on the evolutionary characteristics at He ignition ($L_{\rm tip}$, $L_{\rm ZAHB}$) with respect to other uncertainty sources in evolutionary models (mainly the radiative opacity); furthermore, the obtained variations for the analysed quantities are comparable to the ones presented here. 

\begin{table*}
\begin{center}
\caption{Same as in Table \ref{tab: R_C12}
but for different values of the $3\alpha$ reaction rate.}
\label{tab: R3a}
\begin{tabular}{ lcccccc} 
\hline\hline
& Rate $+34\%$ & Rate $+24\%$ & Rate $+12\%$ & Standard rate & Rate $-12\%$ & Rate $-24\%$ \\ 
 \hline
$R$ parameter & 0.9285 & 0.9372 & 0.9425 & 0.9487 & 0.9681 & 0.9709\\ 
& $-2.0\%$ & $-1.2\%$ & $-0.7\%$ & & $+2.0\%$ & $+2.3\%$ \\
 \hline
 $R2$ parameter & 0.1293 & 0.1289  & 0.1284 & 0.1275 & 0.1258 & 0.1253\\ 
 & $+1.4\%$ & $+1.1\%$ & $+0.7\%$ & & $-1.3\%$ & $-1.7\%$\\
 \hline
\end{tabular}
\end{center}
\end{table*}

As for the $\co$ reaction rate, the effect of changing the $3\alpha$ reaction rate on the theoretical determination of the $R$ and $R2$ parameters was assessed. As reported in Table~\ref{tab: R3a}, the variation in the two parameters is about $\pm 2\%$, which is lower than both the effect due to the uncertainty of the $\co$ rate and the observational errors, making the effect quite negligible.

\section{Conclusions}
\label{sec:conclusions}

In this work, we analyse the effect on the evolutionary characteristics of He-burning stars of the present uncertainty on the $3\alpha$ and $\co$ reaction rates. The $3\alpha$ and $\co$ are competing reactions for central and shell He-burning; thus, they affect several stellar characteristics in these phases. Particularly, they are known to influence the He-burning lifetimes and, more significantly, the carbon and oxygen abundance profile, which, in turn, influences the subsequent WD evolution. Because the uncertainties in these two reaction rates have been considerably reduced over the years, it is important to quantitatively evaluate the present scenario.

We adopted as reference for the $\co$ reaction rate the value proposed by \citet{2017deBoer}, with a quoted uncertainty of $20\%$ at He-burning temperatures, and for the $3\alpha$ the rate estimate by \citet{Fynbo2005}, with a nominal $12\%$ error. Stellar models were computed adopting a $3\alpha$ reaction rate changed up to about $\pm 2 \sigma$; we also considered a value of $+34\%$, from the recent measurements by \citet{2020HoyleKibedi}. The $\co$ rate was modified up to $\pm 35\%$, arising from the estimate by \citet{2002Kunz}.

The two reaction rates were varied independently to study how they affect the evolution of three stellar models: a low mass star ($0.67$~$M_{\sun}$, $Y=0.246$, $Z=0.0001$) with a $0.8$~$M_{\sun}$ as a progenitor, (representative of halo stars) and two intermediate mass models, $1.5$~$M_{\sun}$ and $2.5$~$M_{\sun}$ ($Y=0.28$, $Z=0.015$), with a chemical composition that is typical of disk stars. We also explicitly verified that the uncertainty related to the two reactions propagate independently of the stellar evolution.

 Regarding the $\co$ reaction rate, the present uncertainties marginally affect most of the analysed evolutionary characteristics. The central He-burning lifetime changes by $\sim2\%$ for a rate variation of $\pm20\%$, and by $\sim4\%$ for a change of $\pm 35\%$. The AGB lifetime is affected only at the order of $1\%$. This variation in the evolutionary times affects the theoretical estimate of two important observables for globular clusters, the $R$ and $R2$ parameters, up to $3\%$ and $4\%,$ respectively, for a $\pm 35\%$ change of the reaction rate. On the other hand, the star's chemical profile up to the first thermal pulse is still significantly affected. The $^{12}{\rm C}$ central abundance is modified between $13\%$ and $18\%$, depending on the mass of the star, for a variation of the $\co$ reaction rate by $\pm 1\sigma$. Similarly, the $^{16}{\rm O}$ central abundance is affected at the order of $10\%$ for the same $\co$ rate change ($\pm20\%$). 

The present uncertainty of the $3\alpha$ reaction rate mainly influences the carbon and oxygen central abundances for all the considered models and the AGB lifetime, $t_{\rm AGB}$, for intermediate mass stars. For low-mass stars, the quantities at He ignition, namely, the He core mass, the luminosity of the RGB tip, and the luminosity of the ZAHB are all affected for less than $1\%$. In the case of the $0.67$ $M_{\sun}$ and $1.5$ $M_{\sun}$, the central He-burning lifetime and the AGB lifetime vary less than $1\%$ and $2\%$, respectively, for a $1\sigma$ variation in the reaction rate; on the other hand a change in the $3\alpha$ reaction rate produces a noticeable effect on the $t_{\rm AGB}$ of the $2.5$ $M_{\sun}$. The carbon and oxygen central abundances at the first thermal pulse change between $5\%$ and $6\%$ for $^{12}{\rm C}$, and between $3\%$ and $4\%$ for $^{16}{\rm O}$, when the $3\alpha$ is varied by $1\sigma$. The extreme rate variation of $+34\%$  affects the central carbon abundance by $+15\%$ and the oxygen one by $-9\%$. 

It is worth noting that different methods of treating convective mixing during the central He-burning phase are adopted in the literature. They influence the evolutionary characteristics of stars, particularly the evolutionary times. As reported, for instance, by \citet{CostantinoConvezione1} and \citet{2003Straniero}, depending on the mixing scheme the HB evolutionary time can change by about $4\%$, making the effect more significant than the one due to both the $3\alpha$ and $\co$ reaction rate uncertainties, if the present errors are considered. 

On the other hand, the uncertainty of the chemical profile related to the treatment of convection is still comparable with the one due to the errors of the two reaction rates.
The chemical profile at the end of the shell He-burning phase influences the WD cooling sequence, particularly the cooling time, and WD g-modes
pulsation periods \citep{Degeronimo2017, Chidester2022}. As an example, a change in the $\co$ reaction rate of about $\pm35\%$ can affect the WD cooling time between $7\%$ and $3\%$ depending on the numerical prescriptions adopted to calculate the evolution; this is an effect that is still relevant with respect to other uncertainty sources in this phase, \citep[see e.g.][]{2010SalarisWD}.
Further investigations are needed to analyse in detail the effects of the present uncertainties of the $3\alpha$ and $\co$ rates on the WD evolution of low- and intermediate-mass stars. 

\begin{acknowledgements}
G.V., P.G.P.M., and S.D. acknowledge INFN (Iniziativa specifica TAsP).
\end{acknowledgements}

\bibliographystyle{aa}
\bibliography{bibio}

\appendix

\section{Joint effect of $3 \alpha$ and $\co$ reaction rates}\label{app:linear}

This appendix presents some results obtained by varying simultaneously the two rates at their upper (label hh in Table~\ref{tab:co+3a}) and lower (label ll) boundary.
Thus, hh corresponds to a $\co$ rate of +35\% and $3\alpha$ rate of +24\%. 
From the results by \citet{2013Valle}, we expect that in a small perturbation regime, there is no interaction between the different reaction rates. This is confirmed by the results shown in Table~\ref{tab:co+3a}, which presents the values $t_{\rm HB}$, $t_{\rm AGB}$ as well as the abundances of C and O core at the central He exhaustion. The percent difference, $d,$ with respect of the reference value and the expected value, $d_l$, assuming that the effects of the two rates can be added linearly, are presented. It is apparent that the linearity assumption is justified.

\begin{table}[h]
\begin{center}
  \caption{Joint effect of the $3\alpha$ and $\co$ reaction rates. The columns labelled ll correspond to computation performed with both rates at their lower boundary; those labelled hh to the rates at their upper boundary; $d$ is the percent difference with respect to the reference value, $d_l$ is the effect computed by assuming independence between the two rates. }
\label{tab:co+3a}
\begin{tabular}{lcccccccc} 
\hline\hline  
  & \multicolumn{8}{c}{0.67 $M_{\sun}$}\\
  & \multicolumn{2}{c}{$t_{\rm HB}$ [Myr]} &  \multicolumn{2}{c}{$t_{\rm AGB}$ [Myr] }   &   \multicolumn{2}{c}{$X_{^{12}{\rm C}}$} &   \multicolumn{2}{c}{$X_{^{16}{\rm O}}$}\\
  \hline
 & ll & hh &  ll & hh & ll      & hh & ll & hh\\
 &      95.2    & 100.8 & 12.48 & 12.47 & 0.4627 & 0.3355 & 0.5372 & 0.6644\\
 $d$ &  -3.3\% &        2.4\%    & -0.32\% &    -0.40\% & 20\% & -13\% & -13\%   & 7.9\%\\
 $d_l$  & -3.3\% & 2.4\% & -0.32\% & -0.48\% & 20\% & -13\%     & -13\% & 8.1\%\\
 \hline
  & \multicolumn{8}{c}{1.5 $M_{\sun}$}\\
   & \multicolumn{2}{c}{$t_{\rm HB}$ [Myr]} &  \multicolumn{2}{c}{$t_{\rm AGB}$ [Myr]}    &   \multicolumn{2}{c}{$X_{^{12}{\rm C}}$} &   \multicolumn{2}{c}{$X_{^{16}{\rm O}}$}\\
  \hline
 & ll & hh &  ll & hh & ll      & hh & ll & hh\\
 &      113.4   & 120.1 & 13.18 & 13.33 & 0.4553 & 0.3276 & 0.5245 & 0.6522 \\
  $d$ & -3.4\% &        2.3\%& 0.08\% & 1.2\%   & 21\% &        -13\%   & -13\% & 8.2\%\\
  $d_l$ & -3.4\% & 2.3\% & 0.08\% &     1.3\% & 20\% & -13\% & -13\% & 7.8\%\\
 \hline 
  & \multicolumn{8}{c}{2.5 $M_{\sun}$}\\
   & \multicolumn{2}{c}{$t_{\rm HB}$ [Myr]} &  \multicolumn{2}{c}{$t_{\rm AGB}$ [Myr]}     &   \multicolumn{2}{c}{$X_{^{12}{\rm C}}$} &   \multicolumn{2}{c}{$X_{^{16}{\rm O}}$}\\
  \hline
 & ll & hh &  ll & hh & ll      & hh & ll & hh\\
& 235.5 & 242.1 & 19.00 & 19.07 & 0.4251 & 0.2991 &     0.5549 &        0.6809\\
$d$  & -2.0\% & 0.79\% &        0.26\% &        0.63\%  & 22\%  & -14\% & -12\% & 7.9\%\\
 $d_l$  & -2.0\% & 0.79\% &     0.21\% &        0.69\% & 22\%   & -14\% & -12\% & 7.9\% \\
 \hline
 \end{tabular}
\end{center}
\end{table}

\end{document}